\documentclass{aastex62}

\graphicspath{{./}{figures/}}

\received{\today}
\revised{\today}
\accepted{?}
\submitjournal{RNAAS}

\shorttitle{New variable young stars in IC\,5070}
\shortauthors{Froebrich et al.}

\begin{document}

\title{Variability in IC\,5070: two young stars with deep recurring eclipses\footnote{Released on June, ?th, 2018}}

\correspondingauthor{Dirk Froebrich}
\email{df@star.kent.ac.uk}

\author{Dirk Froebrich}
\affil{Centre for Astrophysics and Planetary Science, University of Kent, Canterbury, CT2 7NH, UK }

\author{Aleks Scholz}
\affiliation{SUPA, School of Physics \& Astronomy, University of St. Andrews, North Haugh, KY16 9SS, United Kingdom}

\author{Justyn Campbell-White}
\affil{Centre for Astrophysics and Planetary Science, University of Kent, Canterbury, CT2 7NH, UK }

\author{James Crumpton}
\affil{Centre for Astrophysics and Planetary Science, University of Kent, Canterbury, CT2 7NH, UK }

\author{Emma D'Arcy}
\affil{Centre for Astrophysics and Planetary Science, University of Kent, Canterbury, CT2 7NH, UK }

\author{Sally V.~Makin}
\affil{Centre for Astrophysics and Planetary Science, University of Kent, Canterbury, CT2 7NH, UK }

\author{Tarik Zegmott}
\affil{Centre for Astrophysics and Planetary Science, University of Kent, Canterbury, CT2 7NH, UK }

\author{Samuel J.~Billington}
\affil{Centre for Astrophysics and Planetary Science, University of Kent, Canterbury, CT2 7NH, UK }

\author{Ricky Hibbert}
\affil{Centre for Astrophysics and Planetary Science, University of Kent, Canterbury, CT2 7NH, UK }

\author{Robert J.~Newport}
\affil{Functional Materials Group, School of Physical Sciences, University of Kent, Canterbury, CT2 7NH, UK }

\author{Callum R.~Fisher}
\affil{Centre for Astrophysics and Planetary Science, University of Kent, Canterbury, CT2 7NH, UK }


\section{Introduction}

Investigating the structure and properties of the innermost parts of protoplanetary accretion disks on sub-AU scales is currently only possible via indirect methods. One option to map the planet-forming zone is to search for occultations of the central young stellar object (YSO) by circumstellar material, e.g. warps or clumps in the inner disks. Such disk eclipses typically last hours to days \citep{2014AJ....147...82C} and have been identified in massive HAeBe stars such as UX\,Ori \citep{1999AJ....118.1043H} and lower mass objects such as AA\,Tau \citep{1999A&A...349..619B}. Of particular interest are quasi-periodic dimming events. They allow distance determinations of the occulting material from the central star. In such cases the actual azimuthal physical extent of the material can be determined from the duration of the dimming event relative to the period. Observations over several periods enable investigations into temporal changes in the line of sight column density distribution, and multi-wavelength data allows us to probe the dust scattering properties. Our citizen science project HOYS-CAPS \citep{2018MNRAS.tmp.1293F} aims to identify such periodic dimming events around YSOs. We used this data-set to search for periodic signatures in light-curves from YSOs in the Pelican nebula (IC\,5070). For this field we have $\sim$\,200 individual observations in the V, R, and I-band filters, distributed over $\sim$\,800 days. Hence, the average cadence is 4\,days, but the most frequent gap (30\,\%) between subsequent observations is 2\,days. Observations are usually taken as 8\,$\times$\,2\,min integrations in all filters to achieve a consistent S/N. 


\section{Results}

\begin{figure}[ht!]
\plotone{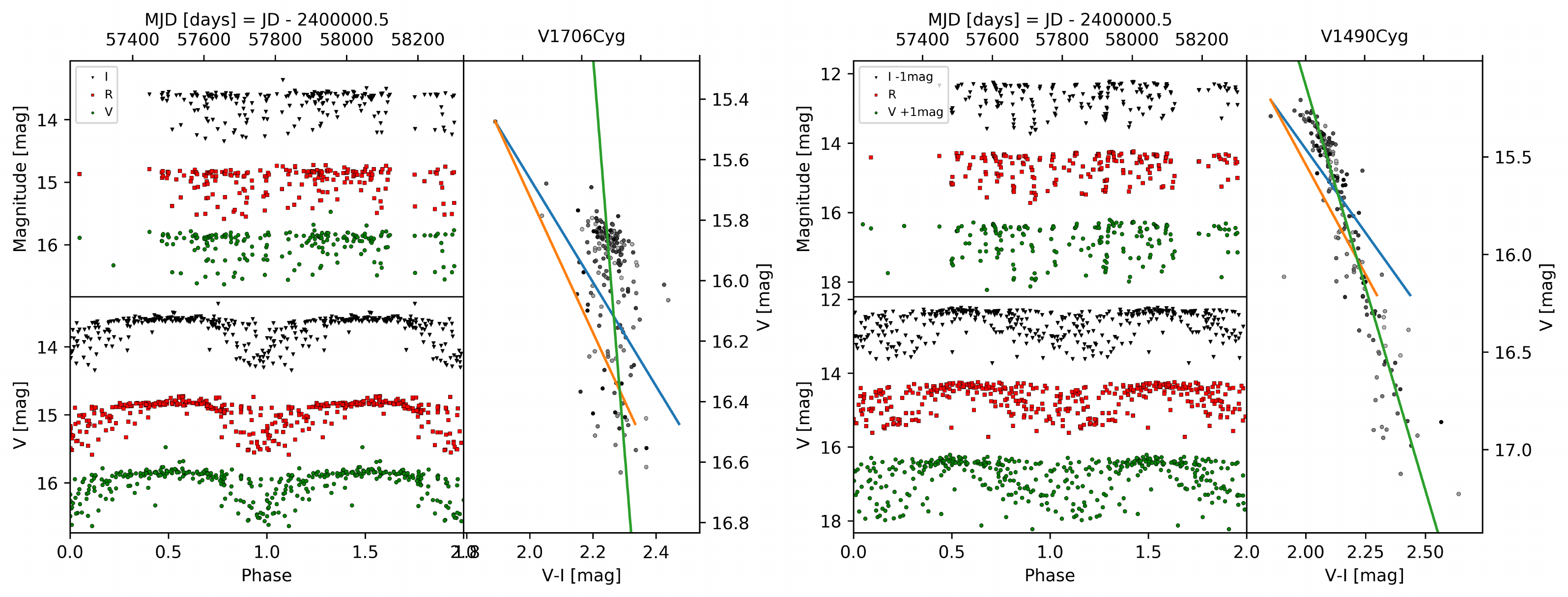} 
\caption{Data for V\,1706\,Cyg (left) and V\,1490\,Cyg (right). In both figures the top left panel shows the HOYS-CAPS V- (green circles), R- (red squares) and I-Band (black triangles) light-curves. In the bottom left panel we show the folded light-curves. At the right is the V vs V-I colour-magnitude diagram. The green line is a fit to the data, while the blue and orange lines indicate 1\,mag optical extinction vectors for R$_V$\,=\,3.1 and 5.0, respectively.  \label{fig}}
\end{figure}

Here we present two newly discovered low-mass YSOs in IC\,5070 with strong variability. Both objects show quasi-periodic dimming events. The colour-magnitude variations of the objects are in agreement with variability caused by changes in the intervening column density of material composed of larger than normal ISM dust or opaque material, indicated by steep reddening vectors in the V vs V-I colour-magnitude diagrams (Fig.\,\ref{fig}). Both sources are known YSOs and are classified as Or$^*$ based on \citet{2003AstL...29..468S}. They are not listed in \citet{2013ApJ...768...93F} who surveyed the area using data from the Palomar Transient Factory\footnote{\tt http://www.astro.caltech.edu/ptf/} nor are they included in the ASAS-SN variable stars database\footnote{\tt https://asas-sn.osu.edu/variables}. 

\noindent{\bf V\,1706\,Cyg, 2MASS\,J20514755$+$4425106:} The object is located at RA\,=\,20:51:47.55 and DEC\,=\,+44:25:10.6 (J2000). The Lomb-Scargle periodogram of the V-band data shows a period of 3.3\,days with an amplitude of 0.8\,mag. The folded light-curve (Fig.\,\ref{fig}) shows that the occultations do not occur with the same depth and duration each time. The object spends up to half the period in a non-bright state. 



\noindent{\bf V\,1490\,Cyg, 2MASS\,J20505357$+$4421008:} The source is located at RA\,=\,20:50:53.58, DEC\,=\,+44:21:00.9 (J2000) and known as an emission-line object \citep{2002AJ....123.2597O}. It was included in the recent study by \citet{2018arXiv180511745I}, who do not report any periodicity. Our analysis indicates a V-Band period of 31.8\,days with an amplitude of up to 1.8\,mag. The occultations change duration and depth and last up to 70\,\% of the period. 



\section{Discussion}

These two newly discovered variable YSOs with quasi-periodic dimming events demonstrate that searching for variable line of sight extinction, using multi-wavelength, ground-based data from small telescopes is a valid approach to study structures in disks. Assuming both stars are IC\,5070 members at about 500\,pc, the out-of-eclipse brightness would indicate stellar masses of 0.5\,--\,1.0\,M$_{\odot}$. The measured periods imply that the occulting disk structures are on a Keplerian orbit with radius 0.034\,--\,0.043\,AU (V\,1706\,Cyg) and 0.15\,--\,0.20\,AU (V\,1490\,Cyg). That means, we are seeing structure in or above the very inner disks.


We can compare the eclipse properties to systematic studies YSO 'dippers' by \citet{2016ApJ...816...69A}, \citet{2018MNRAS.476.2968H} and \citet{2015A&A...577A..11M}. While V\,1706\,Cyg has an eclipse duration typical for such objects, the eclipse depth of about 50\,\% is at the high end of the literature samples. The obscurations in V\,1490\,Cyg are on the very extreme end of the distributions for both duration (up to 22\,days) and depth (up to 80\,\%), and more comparable to UX-Ori type eclipses seen primarily in Herbig Ae/Be stars. Among low-mass T-Tauri stars occultations of that duration and depth have been observed (e.g. \citet{2016MNRAS.463.4459B}), but without periodic recurrence. Thus, these two new discoveries represent extreme examples of disk eclipsing low-mass YSOs and demonstrate that structures in the inner regions of protoplanetary disks can have a wide range of properties.

\begin{thebibliography}{}

\bibitem[Ansdell, et al.(2016)]{2016ApJ...816...69A} Ansdell, M., Gaidos, E., Rappaport, S.~A., et al.\ 2016, \apj, 816, 69.

\bibitem[Bouvier, et al.(1999)]{1999A&A...349..619B} Bouvier, J., Chelli, A., Allain, S., et al.\ 1999, \aap, 349, 619.

\bibitem[Bozhinova, et al.(2016)]{2016MNRAS.463.4459B} Bozhinova, I., Scholz, A., Costigan, G., et al.\ 2016, \mnras, 463, 4459.

\bibitem[Cody, et al.(2014)]{2014AJ....147...82C} Cody, A.~M., Stauffer, J., Baglin, A., et al.\ 2014, \aj, 147, 82.

\bibitem[Findeisen, et al.(2013)]{2013ApJ...768...93F} Findeisen, K., Hillenbrand, L., Ofek, E., et al.\ 2013, \apj, 768, 93.

\bibitem[Froebrich, et al.(2018)]{2018MNRAS.tmp.1293F} Froebrich, D., Campbell-White, J., Scholz, A., et al.\ 2018, \mnras, 1293.


\bibitem[Hedges, et al.(2018)]{2018MNRAS.476.2968H} Hedges, C., Hodgkin, S. \& Kennedy, G.\ 2018, \mnras, 476, 2968.

\bibitem[Herbst \& Shevchenko(1999)]{1999AJ....118.1043H} Herbst, W. \& Shevchenko, V.~S.\ 1999, \aj, 118, 1043.

\bibitem[Ibryamov, et al.(2018)]{2018arXiv180511745I} Ibryamov, S., Semkov, E., Milanov, T., et al.\ 2018, ArXiv e-prints , arXiv:1805.11745.

\bibitem[McGinnis, et al.(2015)]{2015A&A...577A..11M} McGinnis, P.~T., Alencar, S.~H.~P., Guimar{\~a}es, M.~M., et al.\ 2015, \aap, 577, A11.

\bibitem[Ogura, et al.(2002)]{2002AJ....123.2597O} Ogura, K., Sugitani, K. \& Pickles, A.\ 2002, \aj, 123, 2597.

\bibitem[Samus', et al.(2003)]{2003AstL...29..468S} Samus', N.~N., Goranskii, V.~P., Durlevich, O.~V., et al.\ 2003, Astronomy Letters, 29, 468.


\end{thebibliography}
\end{document}